%% file: main.tex
\documentclass[conference, 10pt]{IEEEtran}

\IEEEoverridecommandlockouts
\usepackage[T1]{fontenc}
\usepackage[utf8]{inputenc}
\usepackage{cite}
\usepackage{amsmath, mathtools}
\usepackage{amsfonts,amsthm,bm}
\usepackage{amssymb}
\usepackage{comment}
\usepackage{graphicx}
\usepackage{standalone}
\usepackage{dsfont}
\usepackage{mathtools}
\usepackage{color, colortbl}
\usepackage{soul}
\usepackage[normalem]{ulem}
\usepackage[acronym,shortcuts]{glossaries}
\usepackage{arydshln}
\usepackage{bbm}
\usepackage{algorithm}
\usepackage[noend]{algpseudocode}
\usepackage{adjustbox}
\usepackage[inline]{enumitem}
\usepackage{balance}
\usepackage{tabularx}
\usepackage{booktabs}

\usepackage{comment}

\usepackage{tikz}
\usepackage{pgfplots}
\pgfplotsset{compat=newest}
\pgfplotsset{plot coordinates/math parser=false}
\newlength\fheight
\newlength\fwidth
\usetikzlibrary{plotmarks,patterns,decorations.pathreplacing,backgrounds,calc,arrows,arrows.meta,spy,matrix}
\usepgfplotslibrary{patchplots,groupplots}
\usepackage{tikzscale}

\usetikzlibrary {patterns.meta}

\DeclareMathOperator*{\argmin}{\arg\!\min}

\DeclareMathOperator*{\diag}{diag}
\DeclareMathOperator{\vect}{vec}

\algnewcommand{\IfThenElse}[3]{
  \State \algorithmicif\ #1\ \algorithmicthen\ #2\ \algorithmicelse\ #3}
\algnewcommand{\IfThen}[2]{
  \State \algorithmicif\ #1\ \algorithmicthen\ #2}
\algnewcommand{\LElse}[1]{
  \State \algorithmicelse\ #1}

\newacronym{5g}{5G}{fifth-generation}
\newacronym{6g}{6G}{sixth-generation}
\newacronym{amp}{AMP}{approximate message passing}
\newacronym{awgn}{AWGN}{additive white Gaussian noise}
\newacronym{bs}{BS}{base station}
\newacronym{cpd}{CPD}{canonical polyadic decomposition}
\newacronym{cs}{CS}{compressed sensing}
\newacronym{crc}{CRC}{cyclic redundancy check}
\newacronym{dof}{DoF}{degrees of freedom}
\newacronym{fbl}{FBL}{finite-blocklength}
\newacronym{fec}{FEC}{forward error correction}
\newacronym{lmmse}{LMMSE}{linear minimum mean squared error}
\newacronym{lsfc}{LSFC}{large-scale fading coefficient}
\newacronym{llr}{LLR}{log-likelihood ratio}
\newacronym{mer}{MER}{message error rate}
\newacronym{mimo}{MIMO}{multiple input-multiple output}
\newacronym{ml}{ML}{maximum likelihood}
\newacronym{mmimo}{mMIMO}{massive multiple input-multiple output}
\newacronym{mmtc}{mMTC}{massive machine-type communications}
\newacronym{mra}{MRA}{massive random access}
\newacronym{mrc}{MRC}{maximum-ratio combining}
\newacronym{mtc}{MTC}{machine-type communications}
\newacronym{nlos}{NLos}{non-line-of-sight}
\newacronym{noma}{NOMA}{non-orthogonal multiple access}
\newacronym{pdf}{PDF}{probability density function}
\newacronym{qam}{QAM}{quadrature amplitude modulation}
\newacronym{ra}{RA}{random access}
\newacronym{saloha}{S-ALOHA}{slotted-ALOHA}
\newacronym{scl}{SCL}{successive cancellation list}
\newacronym{sic}{SIC}{successive interference cancellation}
\newacronym{simo}{SIMO}{single input-multiple output}
\newacronym{siso}{SISO}{single input-single output}
\newacronym{snr}{SNR}{signal-to-noise ratio}
\newacronym{svd}{SVD}{singular value decomposition}
\newacronym{tbm}{TBM}{tensor-based modulation}
\newacronym{tbmc}{TBMC}{TBM plus Coherent modulation}
\newacronym{umi}{Umi}{Urban-micro}
\newacronym{ura}{URA}{unsourced random access}
\newacronym{urllc}{URLLC}{ultra-reliable low-latency communications}

\title{Unsourced Random Access With\\Tensor-Based and Coherent Modulations}
\author{Alberto Rech\IEEEauthorrefmark{1}\IEEEauthorrefmark{2}, Alexis Decurninge\IEEEauthorrefmark{1}, and Luis G. Ord\'o\~nez\IEEEauthorrefmark{1}\\ 
\IEEEauthorrefmark{1} Mathematical and Algorithmic Sciences Lab, Huawei France R\&D, Paris, France.\\
\IEEEauthorrefmark{2} Department of Information Engineering, University of Padova, Italy.\\
\small 
\texttt{rechalbert@dei.unipd.it, alexis.decurninge@huawei.com, luis.ordonez@huawei.com}}

\begin{document}
\maketitle

\begin{abstract}
\Ac{ura} is a particular form of grant-free uncoordinated random access wherein the users' identities are not associated to specific waveforms at the physical layer.
\Ac{tbm} has been recently advocated as a promising technique for \ac{ura} due to its ability to support a large number of active users transmitting simultaneously by exploiting tensor decomposition for user separation. 
We propose a novel \ac{ura} scheme that builds upon \ac{tbm} by splitting the transmit message into two sub-messages. This first part is modulated according to a \ac{tbm} scheme, while the second is encoded using a coherent \ac{noma} modulation. 
At the receiver side, we exploit the advantages of \ac{fec} coding and interference cancellation techniques.
We compare the performances of the introduced scheme with state-of-the-art \ac{ura} schemes under a quasi-static Rayleigh fading model, proving the energy efficiency and the fading robustness of the proposed solution.
\end{abstract}

\begin{picture}(0,0)(0,-350)
\put(0,0){
\put(0,0){\qquad \qquad \quad This paper has been submitted to IEEE for publication. Copyright may change without notice.}}
\end{picture}


\glsresetall

\IEEEpeerreviewmaketitle

\section{Introduction}\label{sec:introduction}
The huge density of wireless devices introduces new technological challenges to provide successful connectivity to this increasingly large number of users. The scenario wherein many transmitters aim at transmitting in the uplink to a common \ac{bs} in an uncoordinated manner is referred to as \ac{mra}.
Due to the sporadic nature of the device packet generation (\textit{activations}), only a random subset of transmitters is active simultaneously which typically targets at transmitting small payloads.

A specific type of \ac{mra}, the \ac{ura} paradigm \cite{Polyanskiy17A}, in which the transmitter's identity is not linked to a specific waveform at the physical layer, has recently emerged. 
Within this framework, a variety of practical strategies have been proposed in the literature.
Some \ac{ura} solutions propose to split the transmission slot into multiple sub-slots in which a sub-message is encoded using an inner compressed sensing code. The sub-messages are then concatenated and further encoded using an outer code \cite{Fengler19Grant,Fengler21SPARCs,Shyianov21Massive}. The decoder then combines an inner decoder solving a compressed sensing problem (using e.g. \ac{amp} algorithms) and an outer decoder. 

A different class of approaches that are usually referred to as \textit{pilot}-based schemes proposes to split the transmission slot into two sub-slots. The first slot contains a pilot sequence encoding a small part of the payload that is used at the decoder to estimate the active users channels while the second slot contains a modulation scheme whose decoder assumes the users' channels as known. Note that, in contrast to typical multiple access schemes, wherein pilots are associated with the user identities, pilot bits in \ac{ura} are part of the payload, thus no additional overhead is required for pilot sequences. To avoid misunderstandings, we will refer in the sequel to such symbols as \textit{unsourced signals}. For example, \cite{Fengler21Pilot} proposed to use an \ac{amp} algorithm to decode the unsourced signal and relies on a \ac{mrc} equalizer and a low rate \ac{fec} for the second sub-slot. On the other hand, Gkagkos et al. proposed FASURA in \cite{Gkagkos22FASURA}, a scheme using a simple energy detector to decode the unsourced signal and a scheme combining random spreading and single-user coding for the second slot. Moreover, the decoder uses interference cancellation in order to improve the decoding performance.

\Ac{tbm} is an \ac{ura} scheme that was initially proposed in \cite{Decurninge21Tensor}. In \ac{tbm}, the transmitted symbols are rank-1 tensors so that the combination of the superimposed signals at the \ac{bs} becomes a tensor whose rank is equal to the number of active transmitting users. More specifically, the coded bits of each user are modulated by using Grassmanian sub-modulations in each tensor dimension of the rank-1 tensor.
The main advantage of the \ac{tbm} decoder is that the user separation can be efficiently carried out using tensor decomposition approaches and without relying on pilot sequences. Unfortunately, under practical constraints, the \ac{tbm} receiver suffers from a performance degradation due to the approximations needed to compute the \ac{llr} of the coded bits when the payload is much larger than the tensor sizes.

On a different setting, multiple access schemes assuming knowledge of the users' channel at the receiver (coherent paradigm) have been widely investigated. Traditional orthogonal multiple access have been shown to be outperformed by \ac{noma} schemes \cite{Saito13Non}, for which the symbols of each user are not allocated to orthogonal resources anymore,  in particular in the context of uplink transmissions with \ac{mimo} receivers. Within this setting, an effort has been made towards a unification of the different \ac{noma} solutions and gains in particular in terms of spectral efficiency with respect to their orthogonal counterparts have been highlighted\cite{Meng21Advanced, Chen18Toward}.

In this paper, we propose a novel \ac{ura} strategy, referred to as \ac{tbmc}, which takes the advantages of \ac{tbm} and the pilot-based schemes by combining a \ac{tbm} unsourced signal with a \ac{noma} coherent signal. This allows to overcome the payload size limitations of the original \ac{tbm} and, thus, extends its applicability to a wider range of scenarios. Numerical simulations confirm that \ac{tbmc} achieves competitive performance with respect to state-of-the-art \ac{ura} schemes and is more robust to large-scale fading effects.

The rest of the paper is structured as follows. In Section~\ref{sec:systemmodel}, we introduce the system model. Section~\ref{sec:tbmc} discusses the encoding and decoding procedures in the proposed \ac{tbmc}, which performances are then discussed in comparison with other \ac{ura} state-of-the-art approaches in Section~\ref{sec:numres}. Finally, Section~\ref{sec:conclusions} draws the main conclusions.

\paragraph* {Notation} Scalars are denoted by italic letters, column vectors and matrices by boldface lowercase and uppercase letters, respectively, and sets are denoted by calligraphic uppercase letters. $|\mathcal{A}|$ denotes the cardinality of set $\mathcal{A}$, $[a]$ indicates the set $\{1, \dots, a\}$. $\bm{A}^{\rm T}$ and $\bm{A}^{\dagger}$ denote the transpose and the conjugate transpose of matrix $\bm{A}$, respectively. $\vect(\cdot)$ denotes the vectorization operator, while $\diag(\bm{a})$ denotes the matrix in which entries in the main diagonal are the elements of vector $\bm{a}$. The operator $\otimes$ denotes the Kronecker product. $\mathbb{P}(\cdot)$ denotes the probability operator, and $\mathbb{E}[\cdot]$ the statistical expectation.

\section{System Model}\label{sec:systemmodel}

We consider $K$ single-antenna devices, out of which only a subset of $K_{\rm a}\ll K$ are active simultaneously and aim at transmitting in the uplink to a common \ac{bs} equipped with $M$ antennas. 
Without loss of generality, we assume that the indices of active devices take values in the set $[K_{\rm a}]$. Each active device $k \in \{1\ldots,K_{\rm a}\}$ transmits a message $\bm{m}_k$ of length $B$ bits.
Furthermore, we adopt a quasi-static Rayleigh fading model with uncorrelated channel coefficients between both users and receiving antennas, which remain constant over the $n$ channel resources used for the simultaneous transmissions.

Let $\bm{x}_k \in \mathbb{C}^n$ be the encoded and modulated transmit signal of user $k$ which is normalized to have power $\|\bm{x}_k\|^2 = n$. Then, the received signal at the \ac{bs} is given by
\begin{equation}\label{eq:channel_model}
    \bm{Y} = \sum_{k=1}^{K_{\rm a}}\bm{x}_k\bm{h}_k^{\rm T}+\bm{Z}=\bm{X}\bm{H}^{\rm T}+\bm{Z}
\end{equation}
where $\bm{Z}\in \mathbb{C}^{n\times M}$ is the \ac{awgn} matrix, whose entries are drawn independently from a circularly symmetric complex Gaussian distribution $\mathcal{CN}(0, \sigma^2_z)$, $\bm{h}_k\in \mathbb{C}^M$ captures the channel coefficients between user $k$ and each antenna at the \ac{bs}, which under the assumption of Rayleigh fading are distributed as $\mathcal{CN}(0,\gamma_k)$, where $\gamma_k$ denotes the \ac{lsfc}. In matrix form, symbols and channel vectors are stacked into matrices $\bm{X}= [\bm{x}_1, \dots, \bm{x}_{K_{\rm a}}]$ and $\bm{H}= [\bm{h}_1, \dots, \bm{h}_{K_{\rm a}}]$.

The performance of \ac{ura} schemes is measured by the \ac{bs} decoding capability. Upon the uplink transmissions, the \ac{bs} produces a list $\hat{\mathcal{L}}=\{\hat{\bm{m}}_k:\, k =1,\ldots, \hat K_{\rm a}\}$ of $\hat K_{\rm a}$ decoded messages, which hopefully matches with the list of transmitted packets $\mathcal{L}$. The misdetection and false alarm probabilities are defined, respectively, as
\begin{equation}
    p_{\rm md} = \mathbb{E}\left[\frac{|\mathcal{L}\setminus\hat{\mathcal{L}}|}{K_{\rm a}}\right], \quad  p_{\rm fa} = \mathbb{E}\left[\frac{|\hat{\mathcal{L}}\setminus\mathcal{L}|}{|\hat{\mathcal{L}}|}\right].
\end{equation} 
In the following, we assume that the number of active users is known at the \ac{bs}.
The goal of the modulation design is to minimize the energy-per-bit to noise ratio $E_{\rm b}/N_0 = \frac{||\bm{x}_k||^2}{B\sigma_z^2}$ such that the constraint $P_{\rm e} := \min\{p_{\rm md} + p_{\rm fa}, 1\}\leq\epsilon$ is satisfied, for given $n$, $K_{\rm a}$, $B$, $M$, and $\epsilon$.

\section{\Acf{tbmc}}\label{sec:tbmc}
In this section, we present \ac{tbmc}, a novel unsourced scheme that results from the concatenation of two signals:
\begin{itemize}
\item a first \textit{unsourced signal} encoded as a \ac{tbm} vector symbol responsible for the channel estimation, and carrying a fraction of the message,
\item a second \textit{coherent signal} encoded and transmitted using a \ac{noma} scheme that carries the remaining fraction of the user message and is decoded after channel estimation.
\end{itemize}
Observe that the above structure is similar to that of pilot-based schemes with the particularity that the unsourced signal is a \ac{tbm} vector symbol.
In the following, we will present in detail the \ac{tbmc} encoder and decoder.

\subsection{\ac{tbmc} Encoder}\label{sec:tbmcencoder}

First, the encoder divides the resource block into a first sub-block of $n_{\rm u}$ resources for the unsourced transmissions and a second sub-block of $n_{\rm c}$ resources for coherent signals transmissions, such that $n=n_{\rm u}+n_{\rm c}$. 
Accordingly, the bit sequence to be transmitted by user $k$ is split into two parts $\bm{m}_k^{\rm u}$ and $\bm{m}_k^{\rm c}$, of length $B_{\rm u}$ and $B_{\rm c}$, respectively (so that $B_{\rm u} + B_{\rm c} = B$). The messages $\bm{m}_k^{\rm u}$ and $\bm{m}_k^{\rm c}$ are then encoded into the symbol vectors $\bm{x}_k^{\rm u}$ and $\bm{x}_k^{\rm c}$, respectively. Finally, the transmitted signal $\bm{x}_k$ is given by the concatenation of the unsourced and coherent signals, i.e.,
\begin{equation}\label{eq:ch_input}
    \bm{x}_k = \begin{bmatrix}\bm{x}^{\rm u}_{k}\\\bm{x}^{\rm c}_{k}\end{bmatrix}\in\mathbb{C}^{n}\quad \text{with} \quad\bm{x}^{\rm u}_{k}\in\mathbb{C}^{n_{\rm u}}, \quad \bm{x}^{\rm c}_{k}\in\mathbb{C}^{n_{\rm c}}.
\end{equation}
Therefore, the received signal in \eqref{eq:channel_model} can be rewritten as the concatenation of the two signals:
\begin{equation}
    \bm{Y} = \begin{bmatrix}\bm{Y}^{\rm u}\\ \bm{Y}^{\rm c}\end{bmatrix} =\sum_{k=1}^{K_{\rm a}}\begin{bmatrix}\bm{x}^{\rm u}_{k}\\\bm{x}^{\rm c}_{k}\end{bmatrix}\bm{h}_k^{\rm T}+\begin{bmatrix}\bm{Z}^{\rm u}\\\bm{Z}^{\rm c}\end{bmatrix},
\end{equation}
where $\bm{Z}^{\rm u}$ and $\bm{Z}^{\rm c}$ are the \ac{awgn} components affecting the unsourced and coherent received signals, respectively.

Let us define $\bm{z}^{\rm u} = \vect(\bm{Z}^{\rm u}) \in \mathbb{C}^{Mn_{\rm u}}$.
For convenience, we reformulate the signal received at the \ac{bs} in the first sub-block with the $n_{\rm u}$ channel resources reserved for the unsourced signals in its vectorized form as 
\begin{equation}
     \bm{y}^{\rm u} = \vect(\bm{Y}^{\rm u})= \sum_{k=1}^{K_{\rm a}}\bm{x}^{\rm u}_{ k}\otimes\bm{h}_k+\bm{z}^{\rm u}.
\end{equation}

On the other hand, the received signal in the second sub-block with the remaining $n_{\rm c}$ resources, can be rearranged in matrix form as
\begin{equation}
     \bm{Y}^{\rm c} = \bm{X}^{\rm c}\bm{H}^{\rm T}+\bm{Z}^{\rm c}. 
\end{equation}

\vspace{5pt}
\noindent\emph{\ac{tbm} Encoder for Unsourced Signals.}
Following the \ac{tbm} scheme introduced in \cite{Decurninge21Tensor}, the unsourced message $\bm{m}_k^{\rm u}$ is firstly encoded with a \ac{fec} code and then modulated with \ac{tbm} as follows. Let us assume that $n_{\rm u}$ can be factorized as $n_{\rm u} = \prod_{i=1}^{D}n_i$ for some $D\geq 2$ and $n_1,\dots,n_D \geq 2$. The transmitted unsourced signal $\bm{x}^{\rm u}_{k}$ is given by the vectorized representation of a rank-1 tensor of dimensions $n_1,\dots,n_D$, that is
\begin{equation}\label{eq:tensor_cost}
\bm{x}^{\rm u}_{k} = \bm{a}_{k,1}\otimes\cdots\otimes \bm{a}_{k, D} \in \mathbb{C}^{n_{\rm u}}
\end{equation}
where each vector $\bm{a}_{k, i}$ belong to a \textit{sub-constellation} denoted by $\mathcal{C}_i$. 
The rank-1 components in \eqref{eq:tensor_cost} can be retrieved at the receiver up to a scalar constant using the \ac{cpd}. \ac{tbm} solves this indeterminacy by constraining the sub-constellations $\mathcal{C}_i,\, i=1,\ldots, D$ to follow a Grassmanian codebook structure. In particular, as the original \ac{tbm} scheme described in \cite{Decurninge21Tensor}, we adopt the so-called cube-split modulation introduced in \cite{Ngo20Cube}. 

\begin{figure}
    \centering
    \includegraphics[width = \columnwidth]{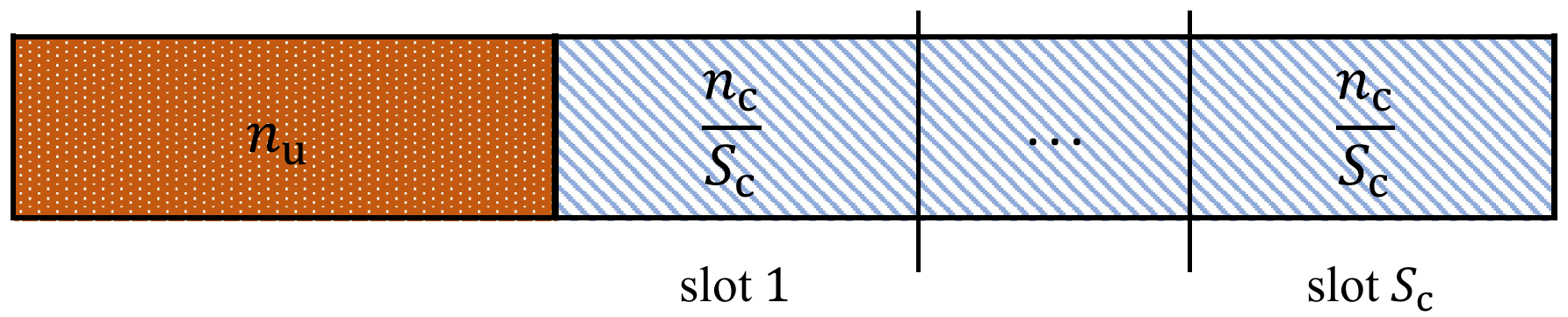}
    \caption{\ac{tbmc} resource block splitting.}
    \label{fig:resources}
\end{figure}

\vspace{5pt}
\noindent\emph{\ac{noma} Encoder for Coherent Signals.}
For the coherent part we adopt the unified \ac{noma} framework introduced in \cite{Meng21Advanced}, wherein data bits are first encoded with a \ac{fec} code and then modulated according to an arbitrary coherent modulation. 
Specifically, we assume that the set of resources for coherent transmissions is split into $S_{\rm c}$ slots of equal size $n_{\rm s} = n_{\rm c}/S_{\rm c}$.
Then, user $k$ selects the slot for transmission based on the unsourced messages as $s_k = \delta(\bm{m}_k^{\rm u})$, 
where $\delta:\{0,1\}^{B_{\rm u}}\rightarrow [S_{\rm c}]$ is a map such that the sets $\delta^{-1}(\{s\})$ are of equal size for all $s=1,\dots,S_{\rm c}$.
Finally, the coherent part of the message $\bm{m}_k^{c}$ is encoded with a \ac{fec} code and modulated using \ac{qam}. Let $\bm{q}_k$ denote the $n_{\rm s}$-dimensional vector collecting the \ac{qam} symbols from user $k$, then the transmitted coherent signal takes the form
\begin{equation}\label{eq:ch_input_data}
    \bm{x}^{\rm c}_{k} =   (\bm{e}_{s_{k}}\otimes\bm{I}_{n_{\rm s}})\bm{q}_k,
\end{equation}
where $\bm{e}_{s_{k}}\in \mathbb{R}^{S_{\rm c}}$ is a vector with $1$ in position $s_k$ and $0$ elsewhere.

A representation of the \ac{tbmc} resource block is depicted in Fig~\ref{fig:resources}. The resources for unsourced signals are orange-colored with a dotted pattern. The blue part with parallel diagonal lines pattern, instead, represents the resources for \ac{noma} coherent transmissions, split in $S_{\rm c}$ slots of equal size.

\subsection{\ac{tbmc} Decoder}\label{tbmcdecoder}
The \ac{bs} decoding task is carried out iteratively, performing a sequence of sub-tasks. First, the received unsourced signal is decoded performing a \ac{cpd}-based user separation (which also allows for the channel estimation), followed by a \ac{llr}-based \ac{fec} decoding and \ac{sic} steps. 
The decoded unsourced signals and the estimated channel vectors are used as input to the \ac{noma} decoder, which performs \ac{lmmse} estimation of the received symbols, as well as \ac{fec} decoding and \ac{sic}.
Finally, the messages are re-combined and re-encoded to obtain the channel inputs and perform a final \ac{sic} step. This procedure is repeated until the decoder either reaches a maximum number of performed iterations or decodes exactly $K_{\rm a}$ messages. 
The iterative \ac{tbmc} decoder is summarized in Algorithm~\ref{alg:bs-decoding}.
Overall, the decoder implements three \ac{sic} iterative procedures: one for the unsourced signals, one for the coherent signals transmissions, and an outer iterative procedure nesting such loops. In the following, we denote the corresponding iterations indices as $j_{\rm u}$, $j_{\rm c}$, and $j$, respectively.
 
\vspace{5pt}
\noindent\emph{\ac{tbm} Decoder for Unsourced Signals.}
When $K_{\rm a}$ is known at the receiver, the joint \ac{ml} multi-user detection and channel estimation problem of the unsourced transmissions is defined as
\begin{equation}\label{mlest}
 \{\hat{\bm{x}}^{\rm u}_{k, i}\} = \argmin_{\substack{\{\bm{x}^{\rm u}_{k,i} \in \mathcal{C}_i\}}} \min_{\substack{\bm{h}_k \in \mathbb{C}^M}}\bigg\lVert\bm{y}^{\rm u}-\sum_{k=1}^{K_{\rm a}}\bm{x}^{\rm u}_{k}\otimes\bm{h}_k\bigg\rVert_2^2.
\end{equation}
Solving \eqref{mlest} directly requires an exhaustive search with $2^{B_{\rm u}K_{\rm a}}$ evaluations of the objective function. Even though the bit sequence length is just a fraction of the message size $B$, the exhaustive search is unfeasible for practical values of $B_{\rm u}$ and $K_{\rm a}$. 
In order to reduce the computational complexity, we split the original \ac{ml} problem \eqref{mlest} into the \textit{user separation} and the \textit{single-user decoding} tasks.
Let us first consider the continuous relaxation of problem \eqref{mlest}:
\begin{equation}\label{mlrelaxed}
    \{\hat{\bm{v}}_{k, i}, \hat{\bm{h}}_k\} = \argmin_{\substack{\bm{v}_{k,i} \in \mathbb{C}^{T_i}\ \\ \bm{h}_k \in \mathbb{C}^M}} \bigg\lVert\bm{y}^{\rm u}-\sum_{k=1}^{K_{\rm a}}\bm{v}_{k,1}\otimes\ldots\otimes\bm{v}_{k,D}\otimes\bm{h}_k\bigg\rVert_2^2
\end{equation}
where $i = 1,\ldots,D$. Problem \eqref{mlrelaxed} can be tackled with an approximate \ac{cpd} algorithm to decompose the $K_{\rm a}$-rank tensor $\bm{y}$ into $K_{\rm a}$ rank-$1$ tensors.
Once the rank-$1$ components are separated, we can perform, independently for each user $k$, single-user decoding of all $\hat{\bm{v}}_{k,i} \in \mathbb{C}^{T_i}$, $i=1,...,D$.
To this end, we use soft demapping by computing the \acp{llr}, which, assuming that the $\ell$-th coded bit of
user $k$, denoted by $b_{k, \ell}$, is among the set of bits mapped into $\bm{a}_{k, i}$, are defined as
\begin{equation}
    {\rm LLR}_{k,\ell} = \log \frac{\mathbb{P}(b_{k,\ell}=1|\hat{\bm{v}}_{k, i})}{\mathbb{P}(b_{k,\ell}=0|\hat{\bm{v}}_{k, i})}.
\end{equation}
An approximated close form for the \ac{llr} computations has been derived in 
\cite{Decurninge21TensorDecomposition} and it is further simplified here as\begin{equation}\label{eq:llrapprox}
    {\rm LLR}_{k,\ell} \approx \left\{\begin{array}{ll}
    \Big( |\hat{\bm{v}}_{k,i}^\dagger\hat{\bm{a}}_{k,i} |  - |\hat{\bm{v}}_{k,i}^\dagger\hat{\bm{a}}_{k,i}^{\text{flip},\ell}|\Big)\hat\eta_{k,i} & \text{if $\hat{b}_{k,\ell} = 1$}\\
    \Big( |\hat{\bm{v}}_{k,i}^\dagger\hat{\bm{a}}_{k,i}^{\text{flip},\ell} |  - |\hat{\bm{v}}_{k,i}^\dagger\hat{\bm{a}}_{k,i}|\Big)\hat\eta_{k,i} & \text{if $\hat{b}_{k,\ell} = 0$}
    \end{array}\right.
\end{equation}
where $\hat{\bm{a}}_{k,i}= \arg\max_{\substack{\bm{a}_{i} \in \mathcal{C}_i}} |\hat{\bm{v}}_{k,i}^\dagger\bm{a}_{i}|$ is the codeword closest to $\hat{\bm{v}}_{k,i}$, $\hat{b}_{k,\ell}$ is the $\ell$-th bit corresponding to $\hat{\bm{a}}_{k,i}$ and $\hat{\bm{a}}_{k,i}^{\text{flip},\ell}$ is the codeword corresponding to the estimated bit sequence with a flipped $\ell$-th bit. Moreover, $\hat\eta_{k,i}$ denotes an approximate variance inverse computed as (see \cite{Decurninge21TensorDecomposition})
\begin{equation}
    \hat\eta_{k,i} = \frac{||\hat{\bm{h}}_k||^2n}{(n_i-1)\sigma_z^2}.
\end{equation}

The \ac{fec} decoder of user $k$ takes as input all the terms ${\rm LLR}_{k,\ell}, 1\leq \ell\leq B_c$ and gives as output the estimated messages.
In general, at iteration $j_{\rm u}$, $\hat{K}_{{\rm a}, j_{\rm u}}\leq K_{\rm a}$ messages are decoded in the $n_{\rm u}$ unsourced signals resources.  
Each decoded bit sequence is subsequently re-encoded and modulated as in \eqref{eq:tensor_cost}. 
If $\hat{K}_{{\rm a}, j_{\rm u}}< K_{\rm a}$ the \ac{sic} step detailed in Section~\ref{sec:sic} is applied to obtain $\bm{y}^{\rm u}_{j_{\rm u}+1}$ and the procedure is repeated while $j_{\rm u}\leq j_{\rm u}^{\rm max}$. 
If, instead, $\hat{K}_{{\rm a}, j_{\rm u}} = K_{\rm a}$, the \ac{tbm} inner decoding loop is stopped, and the estimated channels and messages are passed to the coherent signals decoder.

\vspace{5pt}
\noindent\emph{\ac{noma} Decoder for Coherent Signals}
Upon the decoding of the unsourced messages, the \ac{noma} decoder aims at obtaining the transmitted coherent messages, given the estimated channel vectors $\hat{\bm{h}}_{k}$ and the already decoded unsourced messages $\hat{\bm{m}}^{\rm u}_{k}$, $k = 1, \ldots, \hat{K}_{{\rm a}, j_{\rm u}}$ at the last iteration performed at the \ac{tbm} decoder.
The estimated slots used for the coherent signals transmissions $\hat{s}_k$, $k = 1, \ldots, \hat{K}_{{\rm a}, j_{\rm u}}$ are obtained from the already decoded unsourced messages as $\hat{s}_k = \delta(\hat{\bm{m}}^{\rm u}_k)$,  $k = 1, \ldots, \hat{K}_{{\rm a}, j_{\rm u}}$.
Let $s$ be a generic slot index, and let $\hat{\bm{H}}_s$ be the matrix whose columns are the channel vectors $\hat{\bm{h}}_k$ such that $\hat{s}_k = s$, the \ac{lmmse} filter is applied to retrieve the \ac{qam} symbols transmitted by each active user in each slot $s$ as 
\begin{equation}\label{slot_dec}  
 [\hat{\bm{x}}^{\rm c}_1,\cdots, \hat{\bm{x}}^{\rm c}_{\hat{k_s}}]^{\rm T} = (\hat{\bm{H}}_s^{\dagger}\hat{\bm{H}}_s + \sigma_z^2\bm{I}_{\hat{k_s}})^{-1}\hat{\bm{H}}_s^{\dagger}(\bm{Y}^{\rm c}_s),
\end{equation}
where $\hat{k_s}$ is the estimated number of users transmitting in slot $s$ and $\bm{Y}^{\rm c}_s$ is the signal received in slot $s$. 
Similarly to the \ac{tbm} decoder, the symbols in \eqref{slot_dec} are passed to a \ac{llr}-based \ac{fec} decoder that outputs the estimated coherent messages $\hat{\bm{m}}_k^{\rm c}$, $k=1,\ldots, \hat{K}_{{\rm a}, j_{\rm c}}$, with $\hat{K}_{{\rm a}, j_{\rm c}}\leq\hat{K}_{{\rm a}, j_{\rm u}}$.
Once again, interference cancellation (see Section~\ref{sec:sic}) is applied to obtain $\bm{Y}^{\rm c}_{j_{\rm u} + 1}$ if $\hat{K}_{{\rm a}, j_{\rm c}}<\hat{K}_{{\rm a}, j_{\rm u}}$. The decoding loop is stopped either if the number of decoded unsourced and coherent messages coincides, or the maximum number of iterations $j_{\rm c}^{\rm max}$ is reached. 

\paragraph*{Final Concatenation and \ac{sic}}
At each iteration of the outer loop $j$, upon the decoding of both the unsourced and coherent parts a list of messages $\hat{\bm{m}}_k = [\hat{\bm{m}}_k^{\rm u}, \hat{\bm{m}}_k^{\rm c}]$, for $k =1,\ldots, \hat{K}_{{\rm a},j_{\rm c}}$, is built. If $\hat{K}_{{\rm a}, j_{\rm c}^{max}}<\hat{K}_{{\rm a}, j_{\rm u}}$, one or more coherent signals were not decoded. In this case, the respective unsourced bit messages are dropped. 
As a final step \ac{sic} is applied to the combined received signals to obtain $\bm{Y}_{j+1}$, and the outer loop keeps iterating until $j = j^{\rm max}$ or $K_{\rm a}$ messages are decoded. 

\subsection{Successive Interference Cancellation (SIC)}\label{sec:sic} 
Recall that an interference cancellation step is applied at each iteration of each loop in the decoding process. At given generic iteration $\nu$, a set of $ \hat{K}_{{\rm a},\nu}$ messages are decoded. Those messages are re-encoded with the same \ac{fec} and modulated according to the procedures explained in Section~\ref{sec:tbmcencoder} to obtain the generic channel input $\hat{\bm{X}}$. Then, the channels are estimated with \ac{lmmse} as 
\begin{equation}
    \hat{\bm{H}} = (\hat{\bm{X}}\hat{\bm{X}}^{\dagger} + \sigma_z^2\bm{I}_{M})^{-1}\hat{\bm{X}}^{\dagger}\bm{Y}
\end{equation}
and the interference is mitigated by subtracting the contribution of the users being already decoded as
\begin{equation}\label{eq:sic_data}
    \bm{Y}_{\nu+1} = \bm{Y}_{\nu} - \hat{\bm{X}}\hat{\bm{H}}^{\rm T}.
\end{equation}

\begin{algorithm}[t]
\caption{\ac{tbmc} iterative decoder}
\label{alg:bs-decoding}
\begin{algorithmic}[1]
\Require $\bm{Y}$
\Ensure $\hat{\bm{h}}_k$, $\hat{\bm{x}}_k$, $\hat{\bm{m}}_k$, $\forall k$
\For{$j=1, \ldots, j^{\rm max}$}
\For{$j_{\rm u}=1, \ldots, j_{\rm u}^{\rm max}$}
\State $\{\hat{\bm{v}}_{k, i}, \hat{\bm{h}}_k\} \gets$ \ac{cpd} on $\bm{y}^{\rm u}_{j_{\rm u}}$ to solve \eqref{mlrelaxed}
\State $\hat{\bm{m}}_k^{\rm u}$, $k = 1, \dots, \hat{K}_{{\rm a},j_{\rm u}} \gets$ \ac{fec} decoding on $\{\hat{\bm{v}}_{k, i}\}$
\If{$\hat{K}_{{\rm a},j_{\rm u}}=K_{\rm a}$} 
\State \textbf{STOP} inner loop
\Else \State $\bm{y}^{\rm u}_{j_{\rm u}+1}gets$ \ac{sic} step 
\EndIf
\EndFor
\State compute $\hat{s}_k = \delta(\hat{\bm{m}}^{\rm u}_k)$, for $k=1,\ldots, \hat{K}_{{\rm a}, j}$ 
\For{$j_{\rm c}=1, \ldots, j_{\rm c}^{\rm max}$}
\For{$s=1,\ldots,S_{\rm c}$}
\State $\hat{\bm{x}}_k^{\rm c} \gets$ \ac{lmmse} estimate \eqref{slot_dec} for $k:\: s=s_k$ 
\EndFor 
\State $\hat{\bm{m}}_k^{\rm c}$, $k = 1, \dots, \hat{K}_{{\rm a},j_{\rm c}} \gets$ \ac{fec} decoding on $\hat{\bm{x}}^{\rm c}_k$
\If{$\hat{K}_{{\rm a},j_{\rm c}}=\hat{K}_{{\rm a},j_{\rm u}}$} 
\State \textbf{STOP} inner loop
\Else \State $\bm{Y}^{\rm c}_{j_{\rm c}+1}\gets$ \ac{sic} step 
\EndIf
\EndFor
\State $\hat{\bm{m}}_k\gets[\hat{\bm{m}}_k^{\rm u}, \hat{\bm{m}}_k^{\rm c}]$, for $k =1,\ldots, \hat{K}_{{\rm a},j_{\rm c}}$
\State $\bm{Y}^{\rm c}_{j_{\rm c}+1}\gets$ \ac{sic} step
\EndFor
\end{algorithmic}
\end{algorithm}

\section{Numerical Results}\label{sec:numres}
In this section, we present the numerical results comparing \ac{tbmc} with two state-of-the-art \ac{ura} solutions. 
We assume that the \ac{bs} is equipped with $M=16$ antennas and each active user transmits a message of $B=200$ bits in a block of $n=1000$ channel resources.
In addition to the proposed \ac{tbmc} strategy, we also consider the original \ac{tbm} scheme in \cite{Decurninge21Tensor} using the \ac{tbm} decoder described in Section~\ref{tbmcdecoder}
Note that this corresponds to \ac{tbmc} with $n_{\rm u} = n$ and $B_{\rm u}=B$. For completeness, we further include in our numerical analysis both the pilot-based \cite{Fengler21Pilot} and the FASURA \cite{Gkagkos22FASURA} schemes.

\paragraph*{\ac{fec} Code Details} All messages are encoded with a polar code \cite{Bioglio21Design} comprising \ac{crc}.
At the receiver, the \ac{fec} decoder uses \ac{scl} decoding with a list of candidate codewords of length $L=32$. A \ac{crc} validation step is applied to each codeword in the list and the most likely message among the ones succeeding the \ac{crc} check is returned. If no messages with valid \ac{crc} are found in the list, no messages are outputted.
The coding rate depends on the modulation specifications. On one hand, for the \ac{tbm} unsourced signals, it is adapted to the number of bits modulated with cube-split \cite{Ngo20Cube} in each tensor dimension. On the other hand, for the coherent signals, the code length is constrained by the number of resources per slot $n_{\rm c}/S_{\rm c}$.     
\paragraph*{\ac{ura} Schemes Details} 
For \ac{tbmc} we consider $n_{\rm u}=500$, and $(n_1, n_2, n_3, n_4) = (5,5,5,4)$ with $4$ coded bits-per-dimension.\footnote{Bits-per-dimension refers to the number of coded bits mapped to a point in sub-constellation $\mathcal{C}_i$ in \eqref{eq:tensor_cost}, see \cite{Decurninge21Tensor, Ngo20Cube} for details.}
The number of bits encoded for unsourced signals is $B_{\rm u}=40$. The sub-block of coherent signals resources for \ac{noma} is split into $S_{\rm c} = 2$ slots and the QAM constellation size is set to $4$.
The original \ac{tbm} scheme is configured with $(n_1, n_2)= (40, 25)$ and $4$ coded bits-per-real-dimension. The choice of this configuration was guided by the following remark. As shown in \cite{Decurninge21Tensor}, \ac{tbm} with smaller tensor dimensions are able to support more users, however, we are here constrained by the payload size. The decoding performance drops significantly due to the approximation in the \ac{llr} computation as the sub-constellations become denser. We have observed in numerical results that Grassmannian constellations cannot support more than $4$ coded bits per real dimension.
For the pilot-based scheme proposed in \cite{Fengler21Pilot}, we consider $16$ pilot bits, $360$ pilot resources, a code length of $1280$, and a \ac{scl} list of length $32$.
Finally, the FASURA scheme is configured with $16$ pilot bits, $280$ pilot resources, a code length of $288$, a \ac{scl} list of length $64$, and spreading sequences of length $5$.
The aforementioned \ac{ura} solutions are also compared to the achievability bound derived in \cite{gao23energy}.

\begin{table} 
\centering
\caption{Complexity Comparison}
\label{tab:complexity}
  \small
\begin{tabular}{l|c|c}
\toprule
     & Unsourced & Coherent\\\midrule
Pilot-based & $O(n_{\rm u}M2^{B_{\rm u}})$ & $O(n_{\rm u}MK_{\rm a})$\\
FASURA & $O(K_{\rm a}^3)$ &  $O(T_{\rm spread}K_{\rm a}^3)$\\\midrule
\ac{tbm} & $O(nM + K_{\rm a}^3)$ & $O(1)$ \\
\ac{tbmc} & $O(n_{\rm u}M + K_{\rm a}^3)$ & $O(K_{\rm a}^3/S_{\rm c}^2)$ \\
 \bottomrule
\end{tabular}
\end{table}

\paragraph*{Decoder Complexity Comparison}
In addition to the decoding performance, the decoding complexity plays a crucial role in \ac{ura}, and therefore it has to be discussed. 
All the considered decoding algorithms implement \ac{scl}-based \ac{fec} decoders and, thus, its complexity can be excluded from the comparison.
For tensor-based schemes, the complexity is mainly dominated by the user separation task. The approximate \ac{cpd} for both \ac{tbmc} and \ac{tbm} is here implemented using the Gauss-Newton algorithm with dog-leg trust region \cite{Sorber13Optimization}, which provides a complexity of $O(n_{\rm u}M + K_{\rm a}^3)$ and $O(nM + K_{\rm a}^3)$, respectively. Additionally, \ac{tbmc} computes the \ac{lmmse} estimator for each \ac{noma} slot, adding a complexity of $O(K_{\rm a}^3/S_{\rm c}^2)$.
Instead, the complexity of the pilot-based scheme \cite{Fengler21Pilot} is dominated by the MMV-AMP algorithm, which has complexity $O(n_{\rm u}M2^{B_{\rm u}})$ due to the matrix multiplication between the received symbol and the codebook matrix. Moreover, \ac{lmmse} is applied to estimate the channels and maximum-ratio combining to recover the coherent signals. While the former requires singular value decomposition with a complexity of $O(n_{\rm u}^3)$, the latter is a product of three matrices, having complexity $O(n_{\rm u}MK_{\rm a})$. Note that the complexity of MMV-\ac{amp} can be reduced to $O(n_{\rm u}MB_{\rm u})$ by adopting a DFT codebook.
Finally, for FASURA, the complexity is dominated by the symbol estimator. The decoder performs an \ac{lmmse} for each spreading sequence matrix, leading to a total asymptotic complexity of $O(T_{\rm spread}K_{\rm a}^3)$, where $T_{\rm spread}$ is the number of spreading sequence matrices.

In Table~\ref{tab:complexity}, we summarize the complexity of each of the compared schemes,\footnote{All the compared algorithms require a different, but limited, number of iterations either for performing \ac{sic}, MMV-\ac{amp}, or \ac{cpd}. In the comparison, we neglect this aspect for the sake of conciseness.} distinguishing the complexity of the decoding task for the unsourced and the coherent resources.
On one hand, we observe that the complexity of \ac{tbm}, \ac{tbmc}, and the pilot-based scheme is mainly due to the decoding of the unsourced signals. On the other hand, the FASURA scheme, for which $T_{\rm spread}$ is proportional to the number of coded bits, has higher complexity in the \ac{noma} decoding.

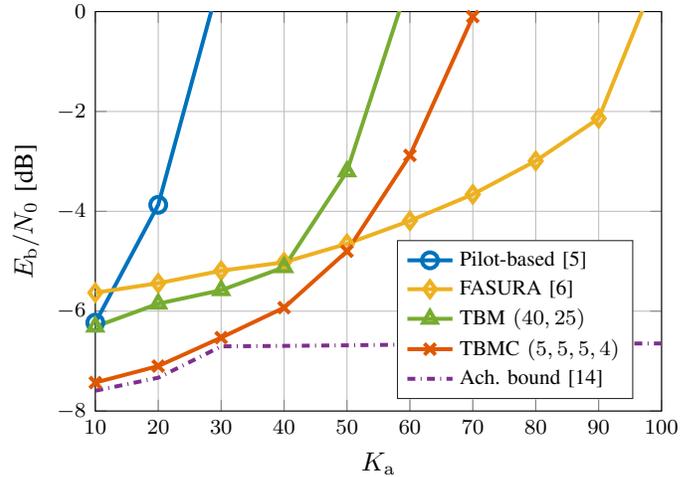
\begin{figure}
    \centering
    \setlength\fwidth{0.85\columnwidth}
    \setlength\fheight{0.6\columnwidth}
    \input{Figs/highSEURA}
    \caption{Average minimum $E_{\rm b}/N_0$ required to achieve $P_{\rm e} \leq 0.1$ versus $K_{\rm a}$, for $n=1000$ and $M = 16$.}
    \label{fig:EbN01000}
\end{figure}

\paragraph*{Energy Efficiency Performance}
The energy efficiency of the \ac{ura} schemes is measured as the minimum $E_{\rm b}/N_0$ required to satisfy the constraint  $P_{\rm e}= \min\{p_{\rm md} + p_{\rm fa}, 1\} \leq \epsilon$.
Fig.~\ref{fig:EbN01000} shows the empirical average minimum $E_{\rm b}/N_0$ required to satisfy this constraint for $\epsilon = 0.1$ as a function of the number of active users $K_{\rm a}$. 
All \acp{lsfc} $\gamma_k$ are set to $1$, for all $k$.
Firstly, we observe that the $(5,5,5,4)$ \ac{tbmc} outperforms \ac{tbm} in all activation conditions, while the $(25, 20)$ is a valid solution only if few users are active simultaneously.
In comparison with FASURA, we observe that the proposed scheme allows greater energy efficiency for low numbers of active users, while FASURA seems to be more efficient in case of high system load.
Indeed, the limitations of tensor-based schemes, when the number of users is large, come from the tensor decomposition that is held in the continuous domain. Gains are therefore expected by taking into account the constellation knowledge. Such study is left as future work. 
However, it is important to highlight that, while the gap between FASURA and the achievability bound of \cite{gao23energy} is large over all the range of considered $K_{\rm a}$ values, \ac{tbmc} approaches the achievability lower bound for $K_{\rm a}<40$, therefore being almost optimal in such conditions.
Finally, we observe that the gap between all the considered schemes and the achievability bound is huge when considering more than $40$ users, suggesting that there is still big room for improvement.

\paragraph*{Fading Robustness}
We now assess the robustness to large-scale fading.
The \acp{lsfc} are derived as
\begin{equation}
    \gamma_k^{\rm [dB]} = \alpha^{\rm [dB]} -\beta\log_{10}(r_k)
\end{equation}
where $r_k$ is the distance between user $k$ and the \ac{bs} in meters, $\beta$ is the path loss exponent, and $\alpha^{\rm [dB]}$ is an additive control term.
In the following, we consider the users to be uniformly distributed in a disk of inner radius $r_{\rm min}$ and outer radius $r_{\rm max}$, thus $r_k$ is distributed as $\mathcal{U}(r_{\rm min}, r_{\rm max})$. The parameter $\alpha^{\rm [dB]}$ is chosen so that the expected received power is equal to $0$~dB, i.e. $\alpha^{\rm [dB]} = \beta\mathbb{E}[\log_{10}(r_k)]$.
Both the path loss exponent and the minimum distance are chosen according to the non-line-of-sight UMi scenario \cite{3gpp.38.901} as $\beta = 35.3$ and $r_{\rm min}=10$.

\begin{figure}
    \centering
       \setlength\fwidth{0.85\columnwidth}
    \setlength\fheight{0.6\columnwidth}
    \input{Figs/fading_comp}
    \caption{Average minimum $E_{\rm b}/N_0$ required to achieve $P_{\rm e} \leq 0.1$ versus $r_{\rm max}$, for $n=1000$ and $M = 16$.}
    \label{fig:fadingcomp}
\end{figure}
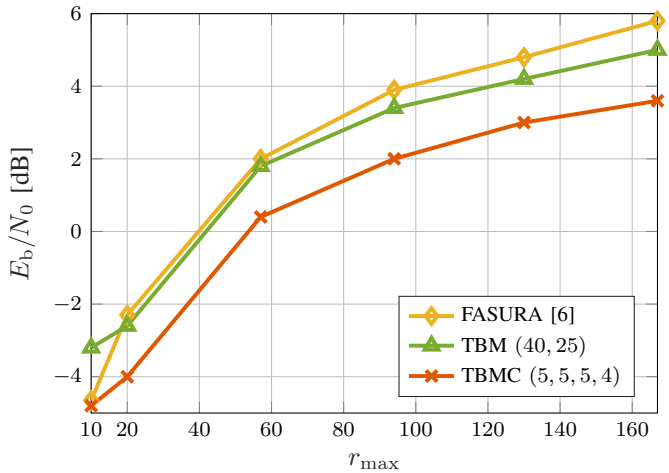

Fig.~\ref{fig:fadingcomp} depicts the minimum $E_{\rm b}/N_0$ guaranteeing $P_{\rm e} \leq 0.1$ as a function of the cell radius, for $K_{\rm a} = 50$ active users and $M=16$ antennas.
The cell radius $r_{\rm max}$ takes values in the range $(r_{\rm min}, 167)$~m, where the upper bound of the interval is the cell radius of a typical UMi cell deployment with an inter-site distance of $250$~m. Note that, the case $r_{\rm max}=r_{\rm min}$ corresponds to the scenario in Fig.~\ref{fig:EbN01000} with $\gamma_k=1$ for all $k$.
Note also that, as the cell radius gets increased, the variance of the received power and therefore the probability of having a relevant fraction of active users exhibiting very poor channel characteristics becomes larger. 
We observe that \ac{tbmc} is the most robust scheme with respect to the received power variance. Instead, despite achieving great performance for $50$ users when the received power is equal for all users (see Fig~\ref{fig:EbN01000}), the FASURA scheme is shown to be more sensitive to the received power variance, as the performance gap with tensor-based schemes increases with the cell radius. This aspect, which should be further investigated, is most likely related to the energy detection phase, which is carried out as the first step at the decoder.

\balance

\section{Conclusions}\label{sec:conclusions}
We presented \ac{tbmc}, an \ac{ura} scheme merging the advantages of \ac{tbm} and pilot-based random access schemes. 
In \ac{tbmc}, the users' payloads and the channel resources are split into two sub-blocks for unsourced and coherent signal transmissions, respectively. For the unsourced signals, \ac{tbm} with state-of-the-art \ac{fec} coding is used to encode a fraction of the message. Then, a \ac{noma} scheme is used for the coherent signal transmissions. At the receiver, soft decoding and \ac{sic} are adopted to mitigate the interference effects.
Numerical results show that with respect to \ac{tbm}, the novel \ac{tbmc} scheme achieves better energy efficiency, approaching the achievability bound for a moderate number of active users. Finally, in a comparison with state-of-the-art solutions, \ac{tbmc} is shown to be more robust to large-scale fading effects.

\bibliographystyle{IEEEtran}
\bibliography{IEEEabrv,Bibliography}

\end{document}

%% file: Figs/highSEURA.tex
\definecolor{mycolor1}{rgb}{0.46600,0.67400,0.18800}%
\definecolor{mycolor2}{rgb}{0.88, 0.34, 0}%
\definecolor{mycolor3}{rgb}{0.00000,0.44700,0.74100}%
\definecolor{mycolor4}{rgb}{0.92900,0.69400,0.12500}%
\definecolor{mycolor5}{rgb}{0.49400,0.18400,0.55600}%

\pgfplotsset{every tick label/.append style={font=\footnotesize}}
\begin{tikzpicture}

\begin{axis}[%
width=\fwidth,
height=\fheight,
at={(0\fwidth,0\fheight)},
scale only axis,
ylabel style={font=\normalsize},
xlabel style={font=\normalsize},
xmin=10,
xmax=100,
xtick={10,  20,  30,  40,  50,  60,  70,  80,  90, 100},
xlabel={$K_{\rm a}$},
ymin=-8,
ymax=0,
ylabel={$E_{\rm b}/N_0$ [dB]},
axis background/.style={fill=white},
xmajorgrids,
ymajorgrids,
legend style={at={(0.95,0.03)}, anchor=south east, legend cell align=left, align=left, font=\footnotesize, draw=white!15!black}]

\addplot [color=mycolor3, line width=1.5pt, mark size=3.0pt, mark=o, mark options={solid, mycolor3}]
  table[row sep=crcr]{%
10	-6.23\\
20	-3.87\\
30  0.73\\
};
\addlegendentry{Pilot-based \cite{Fengler21Pilot}}

\addplot [color=mycolor4, line width=1.5pt, mark size=3.0pt, mark=diamond, mark options={solid, rotate=180, mycolor4}]
  table[row sep=crcr]{%
10	-5.63\\
20	-5.44\\
30	-5.19\\
40	-5.02\\
50	-4.65\\
60	-4.19\\
70	-3.66\\
80	-2.99\\
90	-2.14\\
100 0.97\\
};
\addlegendentry{FASURA \cite{Gkagkos22FASURA}}

\addplot [color=mycolor1, line width=1.5pt, mark size=3.0pt, mark=triangle, mark options={solid, mycolor1}]
  table[row sep=crcr]{%
10	-6.31\\
20	-5.85\\
30	-5.58\\
40	-5.12\\
50	-3.20\\
60	0.66\\
};
\addlegendentry{TBM $(40,25)$}

\addplot [color=mycolor2, line width=1.5pt, mark size=3.0pt, mark=x, mark options={solid, mycolor2}]
  table[row sep=crcr]{%
10	-7.43\\
20	-7.10\\
30	-6.53\\
40	-5.93\\
50	-4.80\\
60	-2.88\\
70  -0.10\\
};
\addlegendentry{TBMC $(5,5,5,4)$}

\addplot [color=mycolor5, dashdotted, line width=1.5pt]
  table[row sep=crcr]{%
10	-7.5936\\
20  -7.3334\\
30	-6.7048\\
40  -6.6953\\
50	-6.6840\\
60  -6.6696\\
70  -6.6571\\
80  -6.6513\\
90  -6.6488\\
100 -6.6469\\
};
\addlegendentry{Ach. bound \cite{gao23energy}}

\end{axis}
\end{tikzpicture}

%% file: Figs/fading_comp.tex
\definecolor{mycolor1}{rgb}{0.46600,0.67400,0.18800}%
\definecolor{mycolor2}{rgb}{0.88, 0.34, 0}%
\definecolor{mycolor3}{rgb}{0.00000,0.44700,0.74100}%
\definecolor{mycolor4}{rgb}{0.92900,0.69400,0.12500}%
\definecolor{mycolor5}{rgb}{0.49400,0.18400,0.55600}%

\pgfplotsset{every tick label/.append style={font=\footnotesize}}
\begin{tikzpicture}

\begin{axis}[%
 width=\fwidth,
height=\fheight,
at={(0\fwidth,0\fheight)},
scale only axis,
xmin=10,
xmax=167,
xlabel style={font=\color{white!15!black}},
xtick={10, 20, 40, 60, 80, 100, 120, 140, 160},
xlabel={$r_{\rm max}$},
ymin=-5,
ymax=6,
ylabel style={font=\color{white!15!black}},
ylabel={$E_{\rm b}/N_0$ [dB]},
axis background/.style={fill=white},
xmajorgrids,
ymajorgrids,
legend style={at={(0.96,0.04)}, anchor=south east, legend cell align=left, align=left, font=\footnotesize, draw=white!15!black}
]

\addplot [color=mycolor4, line width=1.5pt, mark size=3.0pt, mark=diamond, mark options={solid, rotate=180, mycolor4}]
  table[row sep=crcr]{%
10 -4.65\\
20	-2.30\\
57	2.00\\
94	3.90\\
130	4.80\\
167	5.80\\
};
\addlegendentry{FASURA \cite{Gkagkos22FASURA}}

\addplot [color=mycolor1, line width=1.5pt, mark size=3.0pt, mark=triangle, mark options={solid, mycolor1}]
  table[row sep=crcr]{%
10  -3.20\\
20	-2.60\\
57	1.80\\
94	3.40\\
130	4.20\\
167	5.00\\
};
\addlegendentry{TBM $(40,25)$}

\addplot [color=mycolor2, line width=1.5pt, mark size=3.0pt, mark=x, mark options={solid, mycolor2}]
  table[row sep=crcr]{%
10  -4.80\\
20	-4.00\\
57	0.40\\
94	2.00\\
130	3.00\\
167	3.60\\
};
\addlegendentry{TBMC $(5,5,5,4)$}

\end{axis}
\end{tikzpicture}